\documentclass[twocolumn,prb,floatfix]{revtex4}
\usepackage{amsmath}
\usepackage{graphicx}
\usepackage{bm}

\begin{document}

\title{Intramolecular vibrational energy
redistribution from a high frequency mode 
in the presence of an internal rotor: Classical thick-layer diffusion and
quantum localization}

\author{Paranjothy Manikandan and Srihari Keshavamurthy}
\affiliation {Department of Chemistry, Indian Institute of Technology, 
Kanpur, India, 208016}

\begin{abstract}
We study the effect of an internal rotor on the classical and
quantum intramolecular vibrational energy redistribution (IVR) dynamics of 
a model system with three degrees of freedom.
The system is based on a Hamiltonian proposed by Martens and Reinhardt
(J. Chem. Phys. {\bf 93}, 5621 (1990).) to study IVR in the
excited electronic state of para-fluorotoluene.
We explicitly construct the state space and
show, confirming the mechanism proposed by Martens
and Reinhardt, that an 
excited high frequency mode relaxes via
diffusion along a thick layer of chaos created by the low frequency-rotor
interactions.
However, the corresponding quantum dynamics exhibits
no appreciable relaxation of the high frequency mode. 
We attribute the quantum suppression of the 
classical thick-layer diffusion 
to the rotor selection rules and, possibly, dynamical localization effects.
\end{abstract}

\maketitle

Intramolecular vibrational energy redistribution (IVR) has
been the focus of several theoretical and
experimental studies for the past few decades\cite{lsp94,nf96,kmp00,gw04,tu91}.
An important challenge, from the viewpoint of prediction and
control of reaction dynamics, is
to understand the
influence of specific structural features (functional groups)
of a molecule on the IVR dynamics. 
Several 
studies\cite{srh84,mn90,gelps93,hls81,ps86,mp93,hrh83,klps91,bwp94,fpbp93} 
suggest that the early stages of IVR can indeed be
described from such a moiety-specific {\it i.e.,} local 
viewpoint. 
In particular, starting with the experiments by Parmenter and
coworkers\cite{ps86,mpe86,mp93,tpm93}, the role of
large amplitude motions involving internal rotors has
received\cite{wcswj85,fm86,hgrbw88,brz88,sh881,sme88,mr90,zpmbko92,sc92,gelps93,vlf97}, 
and continues to receive\cite{drtm02,bp00,habrw06,pg00,nhc04}, 
considerable attention in
the literature. 

In this work we focus on a mechanism, suggested by
Martens and Reinhardt\cite{mr90} and inspired by the
observations of Parmenter and coworkers\cite{ps86,mpe86},
for the accelarated IVR rates
among the ring modes in para-fluorotoluene (pFT) in comparison to a similar
molecule without the methyl rotor (para-difluorobenzene).
The essential features of the proposed mechanism\cite{mr90}, 
based on detailed classical dynamical studies 
of a model Hamiltonian for pFT
incorporating the methyl rotor,
five of the lowest frequency ring normal modes, 
and important ring-rotor couplings,
can be illustrated using 
the `minimal' Hamiltonian 
\begin{eqnarray}
H&=& \frac{1}{2}\sum_{j=1,2}\left(P_{j}^{2}+\omega_{j}^{2}Q_{j}^{2}\right)+
BP_{\phi}^{2}+\frac{V_{0}}{2}\left[1+\cos 6\phi \right] \nonumber \\
&+& \sum_{j=1,2}\alpha_{j}Q_{j} \sin 3\phi
\label{3dham}
\end{eqnarray}
In the above, three degrees of freedom Hamiltonian,
the ring normal modes $1$, and $2$ 
($\omega_{1}<\omega_{2}$) are coupled with the hindered rotor mode
but the oscillators are not directly coupled to each other.
Setting $\alpha_{1} (\alpha_{2})=0$ one obtains two
degrees of freedom subsystem describing the high (low) frequency mode-rotor
coupling. The subsystem dynamics can be analyzed as follows.
A resonance $\omega_{j}=3 \omega_{r}$, driven
by the coupling terms in Eq.~(\ref{3dham}),
occurs at $P_{\phi}^{(j)} \approx \pm \omega_{j}/6B$ and
leads to quasiperiodic exchange of energy between the oscillator and the rotor. 
However, if in addition the condition $P_{\phi}^{(j)} \sim \sqrt{V_{0}/B}$ 
is satisfied then 
the $\omega_{j}=3\omega_{r}$ resonance can overlap with the
hindered rotor separatrix, $E_{r}=V_{0}$,
generating large scale stochasticity in
the phase space
leading to a rapid relaxation of the oscillator energy. 
Thus one anticipates that if at a given energy
the low frequency-rotor subsystem ($\alpha_{2}=0$, ``bath")
has a strongly chaotic phase space then an excited low frequency mode
will relax rapidly. On the other hand at the same energy 
the high frequency-rotor
subsystem ($\alpha_{1}=0$), typically, would have a near-integrable phase space
and hence an excited high frequency mode will not relax.

\begin{figure}[htbp]
\includegraphics[height=50mm,width=80mm]{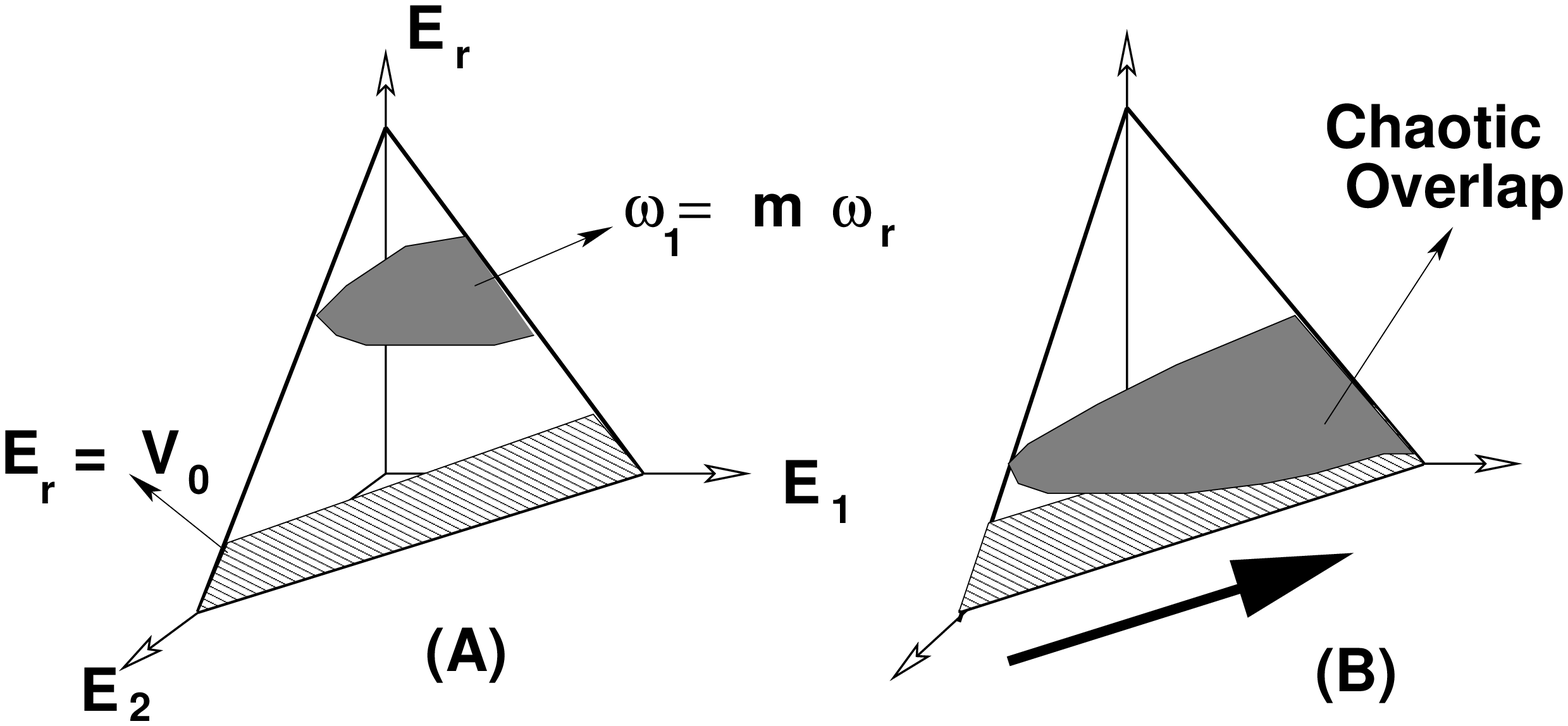}
\caption{Illustrating the mechanism for relaxation of a high
frequency mode ($2$) according to Martens and Reinhardt\cite{mr90}.
In (A) the
frequency $\omega_{1}$ is sufficiently high such that the
$m$:$1$ resonance between the
bath ($1$) and the rotor ($r$)
does not overlap with the hindered rotor separatrix $E_{r}=V_{0}$.
On the other hand, in case (B) the bath frequency is low enough for
an overlap resulting in large scale stochasticity. This band of chaos now
leads to the relaxation of the high frequency system over long time
scales.}
\label{fig1}
\end{figure}

Interestingly, Martens and Reinhardt observed\cite{mr90} that coupling
the subsystems {\it i.e.,} $\alpha_{1},\alpha_{2} \neq 0$ 
leads to a slow relaxation of the excited high frequency mode
despite the lack of resonance overlap. It was suggested\cite{mr90} that
this could be due to a slow
diffusion along the thick layer of chaos created by the bath
subsystem (cf. Fig.~\ref{fig1}B). Evidence for such a ``stochastic pumping"
mechanism\cite{llbook} was provided
by modifying the system so that the bath dynamics is no
longer strongly chaotic as sketched in Fig.~\ref{fig1}A.
Clearly, such a mechanism would lead to extensive mixing of
all the ring modes and therefore support the experimental
observations. 
However, there are crucial questions, given the debate
about the precise role of the rotor\cite{brz88,gelps93,mp93}, 
that have not yet been answered.
Firstly, the Hamiltonian used in the original studies
involved five of the ring modes and the hindered rotor. Is it possible
that the `minimal' Hamiltonian in Eq.~(\ref{3dham}) is enough to
capture the main mechanisms? Secondly, can one explicitly show the
existence of the thick-layer diffusion and understand the nature of
this diffusion? Finally, is the mechanism valid quantum mechanically?
The last issue is crucial, as was also hinted by Martens and Reinhardt,
due to strict rotor selection rules\cite{mpe86,vlf97} and the
possibility that quantum mechanics might
localize the long time classical transport. 
In this work we 
report some of our preliminary results
which provide answers to the questions posed above.

\begin{figure}[htbp]
\includegraphics[height=100mm,width=90mm]{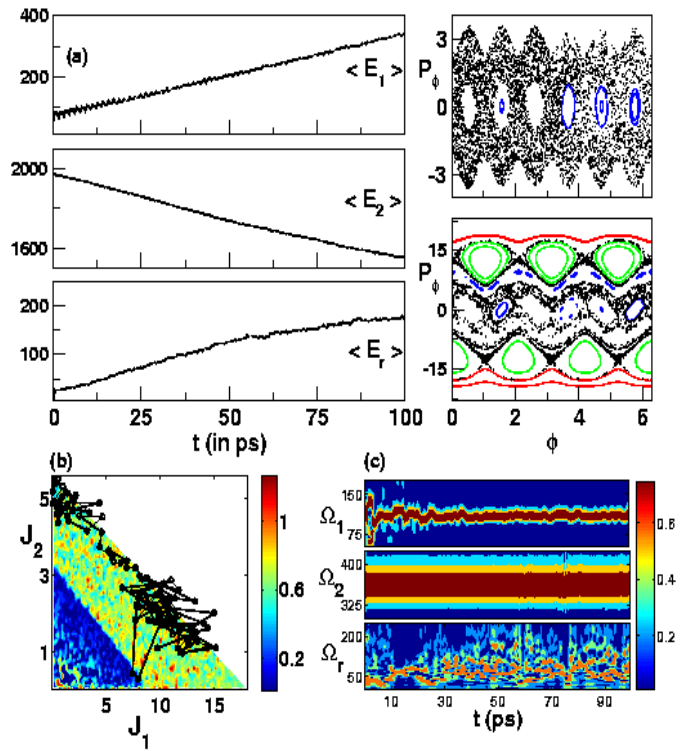}
\caption{(Color online) (a) Classical average mode 
energies (in cm$^{-1}$) versus time for an ensemble of trajectories 
with initial conditions corresponding
to $|v_{1},v_{2},v_{r}\rangle = |0,5,1\rangle$.
The coupling constants\cite{note1} 
are $\alpha_{1}=-3.14$ cm$^{-1}$/$\AA$ and
$\alpha_{2}=2.3$ cm$^{-1}$/$\AA$.
The $(\phi,P_{\phi})$ surface of sections, at the zeroth-order
energy $E_{051}^{0}=2034$ cm$^{-1}$,
for the two degrees of freedom subsystems obtained by decoupling
the high and the low frequency modes respectively are also shown.
The $3$:$1$ resonance islands (green) and the rotor (blue) regions
are clearly visible in the $2$-$r$ subsystem.
(b) The $(\theta_{1},\theta_{2},\phi)=(0,0,0)$ slice of
the state space at $E_{051}^{0}$ is shown and exhibits
regions with large and small diffusions (in ps$^{-1}$). 
Note the existence of a chaotic band which should be compared to the
schematic shown in Fig.~\ref{fig1}B. 
A representative trajectory with large diffusion along the chaotic band
is also shown in (b) with the circles representing $1$ ps
intervals. 
(c) The frequency content of the representative
trajectory in (b) confirms the
chaotic behaviour of the rotor-low frequency subsystem. See text for details.}
\label{fig2}
\end{figure}

We consider the Hamiltonian in Eq.~(\ref{3dham}) with the oscillator
frequencies $\omega_{1}=110$ cm$^{-1}$, $\omega_{2}=359$ cm$^{-1}$,
rotor barrier $V_{0}=34$ cm$^{-1}$, and methyl rotational
constant $B=4.65$ cm$^{-1}$. 
The parameters  
are identical to that of the earlier
work\cite{mr90} except that 
only the lowest and the highest 
frequency modes have been retained\cite{note1}.
An ensemble of classical trajectories with
fixed actions $(J_{1},J_{2},P_{\phi})$ and 
random phases of the oscillators $(\theta_{1},\theta_{2})$ 
and the rotor $(\phi)$
are generated to model initially excited nonstationary
zeroth-order quantum states $|v_{1},v_{2},v_{r}\rangle$.
In Fig.~\ref{fig2} we show the results for 
a representative case of the excited high frequency oscillator
corresponding to the zeroth-order
quantum state $|0,5,1\rangle$.
Trajectories are propagated for $100$ ps and in Fig.~\ref{fig2} the
ensemble averaged mode energies are shown as a function of time.
It is clear that the high frequency mode relaxes over the long timescale
with the energy being gained by the low frequency mode and the rotor.
In fact the subsystem rotor surface of sections shown in Fig.~\ref{fig2}
correspond to $H = E^{0}_{051} \approx 2034$ cm$^{-1}$
and suggest that the high frequency
mode should not relax. Thus the energy flow out of mode $2$ in
Fig.~\ref{fig2}a is a three degree of freedom effect and identical
to the observations made by Martens and Reinhardt in their six
degrees of freedom Hamiltonian\cite{mr90}. 

Is the long time energy relaxation of the mode $2$ seen in 
Fig.~\ref{fig2}a due to the thick-layer diffusion mechanism suggested
earlier\cite{mr90}? To this end we  
study the classical IVR dynamics by explicitly constructing 
the zeroth-order action space $(J_{1},J_{2},P_{\phi})$ {\it i.e.,}
state space. 
A set of initial actions
were chosen by fixing the total energy 
$H=E^{0}_{051}$ and angles $(\theta_{1},\theta_{2},\phi)=(0,0,0)$.
These set of initial conditions are then propagated to $T=100$ ps
and the diffusion constants
\begin{equation}
d_{k}(T) = \frac{1}{T}\int_{0}^{T} \mid \Omega_{k}(t)-\overline{\Omega}\mid dt
\end{equation}
for $k=1,2,r$ are determined. The $\Omega_{k}$ are the 
local nonlinear frequencies extracted using a wavelet-based
time-frequency approach\cite{cwu03}.
The sum of the $d_{k}$'s is 
then associated with each trajectory (point in the state space) and hence
the regular (low diffusion) and chaotic (high diffusion) regions are identified.
In Fig.~\ref{fig2}b the state space is shown as a projection onto the
$(J_{1},J_{2})$ plane and
the band of stochasticity, due to the mode $1$ and rotor
interaction as sketched in Fig.~\ref{fig1}B, is clearly seen.
Several classical trajectories corresponding to $|0,5,1\rangle$ were
propagated for $100$ ps and a variety of diffusive behaviours were seen.
In Fig.~\ref{fig2}b we
show a typical high diffusion trajectory superimposed on the state space
which clearly shows the {\em diffusion along the thick layer of
chaos}. This establishes the mechanism proposed by Martens and Reinhardt
for the minimal Hamiltonian in Eq.~(\ref{3dham}). Moreover, to the best
of our knowledge, this is the first time that the thick-layer diffusion
has been explicitly shown in a molecular system.

\begin{figure}[htbp]
\includegraphics[height=100mm,width=80mm]{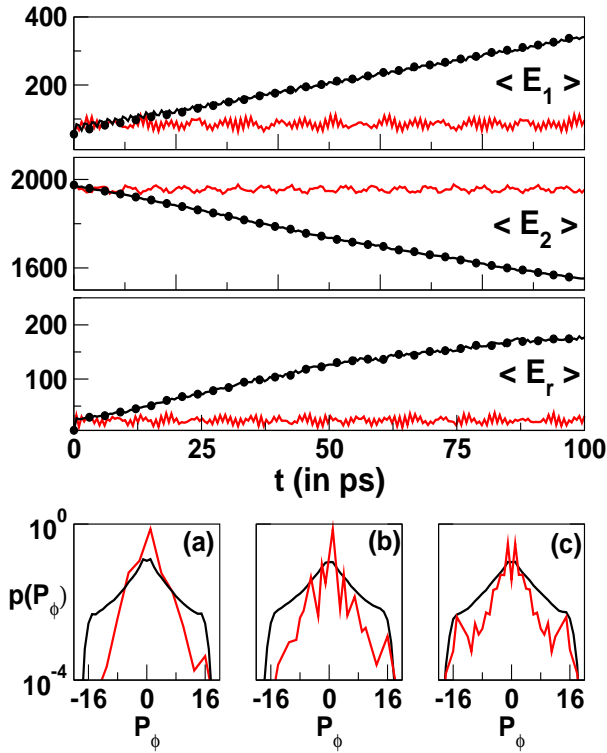}
\caption{(Color online) Quantum average mode
energes (red) compared to the
classical (black symbols) analogs shown in Fig.~\ref{fig2}a.
The bottom panels compare the
classical (thick black) and quantum (red) rotor angular momentum
distributions. (a) Original system (cf. Eq.~\ref{3dham})
indicates localization of
the quantum $p(P_{\phi})$ as compared to the classical case.
In (b) and (c) weak couplings between mode $1$ and the rotor of the form
$\beta Q_{1} \sin 2\phi$ and $\beta Q_{1} \sin \phi$ with
$\beta=-1$ cm$^{-1}$/$\AA$ respectively
are added to the original system in order to break the
$\Delta v_{r}=\pm 3$ selection rule in (a).}
\label{fig3}
\end{figure}

Certain features of the dynamics that are 
inherently due to the system having three or more degrees of freedom
are worth pointing out at this stage.
A careful look at the trajectory shown in Fig.~\ref{fig2}b reveals that
for nearly $15$ ps there is very little
energy relaxation in $E_{2}$. In our computations this feature was common
to a large number of trajectories. 
A preliminary understanding of the $\sim 15$ ps `incubation' time can
be gained by the 
time-frequency analysis\cite{cwu03} of the classical
dynamics. In Fig.~\ref{fig2}c the results for the representative
trajectory undergoing thick-layer diffusion
in the state space are shown as spectrograms {\it i.e.,} the distribution
of the local frequency amplitudes as a function of time. One can
see that during the incubation time the low frequency mode and the rotor
exhibit strong coupling and chaotic dynamics while the high
frequency mode is decoupled. At longer times perturbations in the
local frequency associated with the mode $2$ are clearly visible. 
The rotor phase space shown in Fig.~\ref{fig2} for the bath 
indicates a thin layer of chaos which is insuffcient to
drive the energy relaxation
of the high frequency mode. Consequently we interpret the
incubation time as the time required for the
low frequency-rotor subsystem to generate the thick layer of chaos. 
Incidentally, we remark here that Martens and Reinhardt arrive at 
nearly the same timescale for the high-frequency cutoff in the
spectral density associated with the random force generated
by the chaotic bath dynamics. 
After the incubation time the trajectory
rapidly moves along the chaotic layer resulting in the energy flowing
into the low frequency mode and the rotor. At longer times the trajectory
once again performs diffusive motion about specific regions in
the state space. 

We now address the crucial issue of wether the 
quantum dynamics of Eq.~(\ref{3dham}) would also exhibit the thick-layer
diffusion leading to the relaxation of the excited high frequency mode.
Surprisingly, and as far as we can tell, there has been no attempt
to confirm the mechanism proposed by Martens and Reinhardt in the
quantum domain. We note, however, that Leitner and Wolynes have
studied\cite{lg971,lg972} the effect of quantization 
on the weak stochastic pump model
of Arnol'd diffusion. In this case, involving anharmonic oscillators,
it was found that the quantization
limits the extent of diffusion. Does this imply a similar localization
for the quantum dynamics of our minimal Hamiltonian? To investigate
this possibility we have performed detailed quantum studies of
the IVR dynamics of our system.
A converged basis set, using direct product
of harmonic oscillator and free rotor basis functions,
was employed in all the studies. 
In Fig.~\ref{fig3} we show the quantum average mode energies 
$\langle E_{k} \rangle$ for the zeroth-order state $|0,5,1\rangle$
and compare with the classical results of Fig.~\ref{fig2}.
It is clear from the figure that 
quantum mechanically the high frequency mode
does not relax even on the $100$ ps timescale - infact a beating
pattern is observed. {\em It is important to mention that results of further
studies (not shown here) on
the quantum dynamics including
the rest of the ring modes, as in the original study\cite{mr90}, 
again indicate little to no energy flow out of the high frequency mode}.
Thus, it appears that the classical thick-layer diffusion is completely
suppressed in the corresponding quantum system.

There can be a couple of reasons for the observed quantum behaviour.
The first, as already alluded to earlier, could be due to a genuine
quantum localization of the thick-layer diffusion. Consequently,
relevant eigenstates of Eq.~(\ref{3dham}) were investigated and
none of them displayed a strong delocalization
along the chaotic layer. 
Nevertheless, the fact that the high frequency mode
does not relax even upon the inclusion of additional ring modes
points to an insufficient density of states in the chaotic layer.
One other possibility has
to do with the fact that quantum mechanically there is a strict 
rotor selection rule $\Delta v_{r}=\pm 3$ due to the $Q \sin 3\phi$ 
coupling in Eq.~(\ref{3dham}). Thus the low frequency-rotor coupling
cannot lead to the population of specific rotor states as opposed to
the situation in the classical dynamics. The consequences can
be clearly seen in Fig.~\ref{fig3}a 
where the quantum rotor momentum distribution\cite{rgr89}
\begin{equation}
p(v_{r}=\hbar P_{\phi}) = \frac{1}{T}\int_{0}^{T} dt \sum_{v_{1}v_{2}} 
\left| \langle v_{1}v_{2}v_{r} |e^{-iHt/\hbar} |0,5,1\rangle \right|^{2}
\end{equation}
is compared with the corresponding
classical distribution\cite{mr90} $p(P_{\phi})$ at $T=100$ ps.
The quantum distribution is localized and the
asymmetric form of the envelope
is due to the selection rule. 
In order to assess the importance of the 
rotor selection rule to the observed quantum localization weak
coupling terms between the low frequency mode and the rotor of
the form $\beta Q_{1} \sin 2\phi$ and
$\beta Q_{1} \sin \phi$ were introduced in Eq.~(\ref{3dham}) with 
$\beta=-1$ cm$^{-1}$/$\AA$.
The resulting quantum and classical $p(P_{\phi})$
are compared in Fig.~\ref{fig3}b and Fig.~\ref{fig3}c respectively.
Although one observes population of
higher momentum states, the quantum distribution is still narrower
than the classical distribution. At the same time even for the modified
cases no energy relaxation of the high frequency oscillator is observed.
Therefore, given the lack of a sufficient density of
states in the chaotic layer,
rotor selection rule might not be the sole factor responsible for
the quantum localization. 
The classical-quantum correspondence of $p(P_{\phi})$ have been studied
earlier\cite{rgr89} in the context 
of periodically kicked one degree of freedom systems. In our case the
system has three degrees of freedom and the nature of partial barriers
is not yet clear. Further work, in addition to a careful
study of the dependence of energy diffusion and
quantum eigenstates on $\hbar$,
is required to provide a firm answer and such studies are currently
underway. 
 
The authors acknowledge Profs. Gerrit Groenenboom, 
David Leitner, and Brooks Pate
for useful discussions and suggestions.


\begin{thebibliography}{99}
\bibitem{note1}{The coupling constants $\alpha_{j}$
reported in ref.~\onlinecite{mr90} appear to be too large by a factor of $100$.
This has been confirmed by reproducing the results in
ref.~\onlinecite{mr90} and in this work the correct coupling values
are used.}
\end{thebibliography}
\end{document}